# TÉCNICAS DE ENSEMBLE LEARNING PARA SISTEMA DE DETECÇÃO DE INTRUSÃO NO CONTEXTO DA CIBERSEGURANÇA

Andricson Abeline Moreira
*Department Of Computing,UNESP*
*Universidade Estadual Paulista, Street, São José do Rio Preto, Brasil*

Carlos A. C. Tojeiro
*Department Of Computing,UNESP*
*Universidade Estadual Paulista, Bauru, Brasil*

Carlos J. Reis
*Department Of Computing,UNESP*
*Universidade Estadual Paulista, Bauru, Brasil*

Gustavo Henrique Massaro
*Department Of Computing,UNESP*
*Universidade Estadual Paulista, Street, São José do Rio Preto, Brasil*

Igor Andrade Brito
*Department Of Computing,UNESP*
*Universidade Estadual Paulista, Street, São José do Rio Preto, Brasil*

Kelton A. P. da Costa
*Department Of Computing,UNESP*
*Universidade Estadual Paulista, Bauru, Brasil*

**RESUMO**

Recentemente, houve um interesse em melhorar os recursos disponíveis em técnicas *de Intrusion Detection System* (IDS). Neste sentido, diversos estudos relacionados à cibersegurança demonstram que as invasões de ambiente e o sequestro de informações, são cada vez mais recorrentes e complexas. A criticidade dos negócios envolvendo operações em ambiente utilizando recursos computacionais, não permitem a vulnerabilidade da informação. A cibersegurança tomou uma dimensão dentro do universo da tecnologia indispensável nas corporações, a prevenção de riscos de invasões ao ambiente é tratada diariamente por equipes de Segurança. Assim, o objetivo principal do estudo foi investigar a técnica de *Ensemble Learning* utilizando o método *Stacking*, apoiado pelos algoritmos *Support Vector Machine* (SVM) e *k-Nearest Neighbour* (kNN*)* visando uma otimização dos resultados para detecção de ataques DDoS. Para isso, utilizou-se o conceito de *Intrusion Detection System* com aplicação da ferramenta de *Data Mining* e *Machine Learning* Orange, para obter melhores resultados.

**PALAVRA CHAVE**

Security, Networking, Ensemble, Machine Learning, Intrusion.

## 1. INTRODUÇÃO

Com o passar dos anos observou-se, que em cenários de redes de computadores é imprescindível o uso de ferramentas para detecção de intrusão e de classificação de ataques. Desta forma, várias abordagens tem surgido como propostas de solução para o problema de proteger as redes contra-ataques internos e externos.

'

Porém isto ainda é um grande desafio, em consequência do aumento do tráfego de redes, do surgimento de novos tipos de ataques e da necessidade de centralização de tarefas provindas da web [Ennaji et al., 2021].

Os Sistemas de Detecção de Intrusão (IDSs) são ferramentas de extrema necessidade para segurança, prevenção, e identificação de ataques a dados confidenciais e ameaças originadas na rede. Um IDS possui a função de coletar dados de dispositivos conectados a uma rede, projetado para identificar atividades maliciosas ou ataques contra sistemas [Othman et al., 2018].

Atualmente estes mecanismos de detecção de intrusão, têm como objetivo realizar varreduras detectando invasões, que surgem por meio de atividades suspeitas na rede. Estas ferramentas são voltadas apenas para analisar a rede, onde por meio de seus parâmetros possam identificar invasões com as varreduras e a partir da emissão de alertas, com a funcionalidade de detecção de anomalias e atividades ilegais na rede [Turcato, 2020].

Neste pensamento e diante deste aumento na quantidade de dados que percorrem as redes, fez-se com que os alertas gerados pelos IDS, fossem devidamente classificados com algoritmos de *Machine Learning*, visto que uma das preocupações é a otimização da identificação de taxas de falsos positivos e falsos negativos.

O objetivo principal do estudo é investigar a técnica de *Ensemble Learning* utilizando o método *Stacking*, apoiados pelos algoritmos *Support Vector Machine* (SVM) e k-*Nearest Neighbour* (kNN), aplicados para identificação de ataques de negação distribuída de serviço (DDoS), buscando uma otimização nos resultados.

Dentro deste contexto, apresentamos a proposta da utilização da ferramenta Orange, com os algoritmos kNN e SVM otimizadas com o *Stacking*. O *dataset* usado neste trabalho para os treinos e testes foi o CICDDoS2019, e as métricas de verificação do modelo foi a Acurácia, Precisão, Fl-Score, Recall e Curva Roc.

## 2. TRABALHOS RELACIONADOS

Com base em estudos anteriores notou-se que existem diversas abordagens para detecção de ataques com uso de técnicas tradicionais baseadas em regras [Guan e Ge, 2018] e também com técnicas de *Machine Learning* [Buczak e Guven, 2016].

Em [Arshi et al., 2020], os autores utilizaram técnicas de Aprendizado de Máquina para identificação de ataques DDoS. Os tipos de ataques empregados no estudo foram: UDP Flood, ICMP (PING) Flood, Smurf Attack e HTTP Flood, e os algoritmos utilizados para classificação foram o *Naive Bayes*, SVM, Árvores de Decisão (*Decision Tree*), Rede Neural e K-means.

No trabalho desenvolvido por [Saini et al, 2020], os autores aplicaram técnicas de *Machine Learning* para detecção de ataques DDoS com os classificadores: *Naive Bayes, Random Forest*, *Multi-layer Perceptron* (MLP) e J48. O conjunto de dados utilizado para teino e testes no trabalho, é constituido de 27 atributos e dados de 4 tipos de ataques: Smurf, UDP Flood, SIDDoS e HTTP Flood. Segundo os autores, a taxa de acurácia alcançada foi de 98,64% com o uso do algoritmo J48, 98,63% para MLP, 98,10% para *Random Forest* (RF) e 96,93% para *Naive Bayes*.

A prosposta de [Rajagopal et al., 2020], propõem a abordagem da técnica de *Ensemble Stacking* com os classificadores RF, Regressão Logística (LR) e kNN para detecção de intrusão nos conjuntos de dados UNSW NB-15 e UGR'16. Os autores conseguiram bons resultados na identificação de ataques como DDoS, DoS e identificação de tentativas de scan.

O trabalho de [Chand et al., 2016] tem como propósito principal o uso de SVM com outros classificadores como BayesNet, AdaBoost, IBK, J48, Random Forest, JRip, OneR e SimpleCart juntamente com a técnica de *Stacking*. Os autores conseguiram bons resultados de SVM e Random Forest aplicando Ensemble *Stacking*, atingindo uma taxa de Acurácia de 97,50% nos testes no conjunto de dados NSL-KDD'99.

Em [Yadav et al., 2016], os autores apresentam uma perspectiva em conjunto para a detecção ataques de DDoS utilizando um conceito de *Deep Learning* juntamente com a técnica de *Stacking*. Como resultado concluem que o modelo proposto por eles ajudam a melhorar as taxas de detecção do classificador de regressão logística, atingindo uma taxa média de detecção de 98,99% e uma taxa média de 1,27% de falsos positivos.

## 3. TÉCNICAS UTILIZADAS

As subseções a seguir apresentam as técnicas utilizadas para desenvolvimento do trabalho, bem como, o conjunto de dados, os classificadores, a técnica de Ensemble e a ferramenta de *Data Mining* para o trabalho.

'

## 3.1 Conjunto de Dados CICDDoS2019

O Conjunto de Dados utilizado para o trabalho foi o ClCDDoS2019 [DDoS *Evaluation Dataset* (CIC-DDoS2019), 2019] criado pela *University of New Brunswick*, no Canadá. O ClCDDoS2019 abrange os mais recentes e comuns ataques DDoS como: PortMap, NetBIOS, LDAP, MSSQL, UDP, UDP-Lag, SYN, NTP, DNS e SNMP. Durante o período de desenvolvimento do *dataset* foram feitos vários ataques subsequentes, sendo 12 no primeiro dia de treinamento (NTP, DNS, LDAP, MSSQL, NetBIOS, SNMP, SSDP, UDP, UDP-Lag, WebDDoS, SYN e TFTP) e 7 no dia de testes (PortScan, NetBIOS, LDAP, MSSQL, UDP, UDP-Lag e SYN). Os dados foram salvos em um arquivo com a extensão .csv e incluem informações de rede e logs dos eventos dos ataques.

Conforme [Sharafaldi'n et al., 2019], que propôs uma taxonomia para os ataques DDoS, dividindo-os em ataques de reflexão e ataques de exploração, comentam que os de reflexão são aqueles que tentam fazer com que o alvo responda a sua própria requisição, permitindo ao atacante acesso autenticado pela vítima, podendo ser utilizados em aplicações com os protocolos TCP e UDP ou em ambos.

Já os de exploração são aqueles ataques que ocorrem quando um atacante, aproveitando de uma vulnerabilidade, executam ações maliciosas, como invadir um sistema para acessar informações confidenciais ou disparar ataques contra outros computadores.

Para testar os resultados, foi utilizado o conjunto de dados DDoS NTP (*Network Time Protocol*) parte do *dataset* CICDDoS2019, este conjunto de dados baseado em reflexão UDP possui como objetivo, o envio de várias consultas de curta durações ao servidor, resultando em respostas longas onde são forjados os endereços de retorno das respostas, causando um sobrecarregamento da máquina. A taxonomia proposta está representada na Figura 1.

Figura 1: Taxonomia dos ataques DDoS

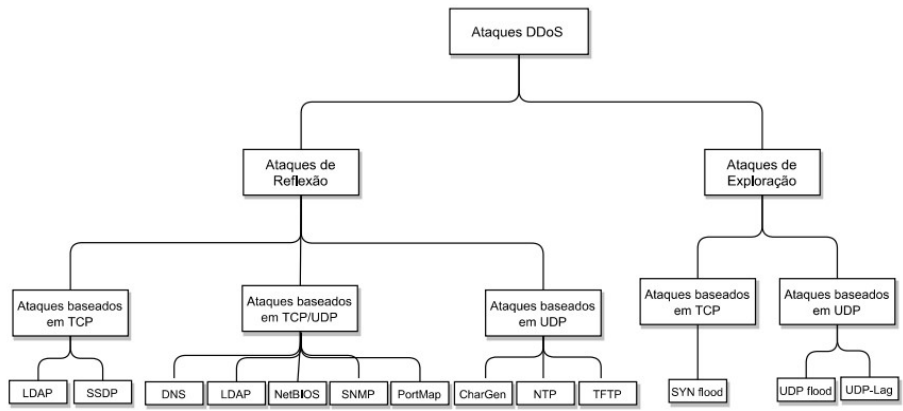

## 3.2 Support Vector Machine (SVM)

O SVM é um algoritmo de Aprendizado de Máquina supervisionado utilizado para classificação e regressão. Pertencem a um grupo de classificadores lineares, usados na classificação de um conjunto de pontos que busca identificar a linha de separação entre duas classes. O SVM possui como função identificar uma reta entre as classes, que pode ser chamado de Hiperplano.

Um modelo SVM é uma representação onde pontos no espaço são mapeados de maneira que os exemplos de cada classe sejam divididos por uma lacuna clara que seja a mais ampla possível, consistindo na seleção de um hiperplano que minimize o risco estrutural, a partir da resolução de um problema convexo quadrático [Cortes et al., 2021].

## 3.3 k-Nearest Neighbour (kNN)

'

O algoritmo supervisionado kNN porposto por Cover e Hart [Cover et al.,1967] é um dos métodos de classificação mais comuns em Aprendizado de Máquina. Possui uma implementação fácil e um desempenho significativo.

O algoritmo kNN, baseia-se na utilização da distância euclidiana para classificar e agrupar itens semelhantes. Cada vez que uma nova instância é lançada ao algoritmo, devem ser calculados os k vizinhos mais próximos da instância a ser classificada [Faria, 2016].

Uma grande desvantagem do algoritmo kNN é com relação a sua escalabilidade pois, é perceptível que o seu processamento fica mais lento à medida que o número de amostras ou a quantidade de dados aumentam.

### 3.4 Método Stacking

O método *Stacking*, também chamado de *Stacked generalization*, foi proposto pela primeira vez por Worlpert [Worlpert, 1992], é uma das técnicas utilizadas pelo método *Ensemble Learning*. Sua função é combinar modelos de classificação ou regressão por meio de um meta-classificador ou meta-regressor.

Inicialmente o conjunto de dados é treinado através de algoritmos básicos de *Machining Learning*, as saídas previamente treinadas nos algoritmos de aprendizados transformam-se em um novo conjunto de dados, que serão utilizadas como entradas para o modelo *Stacked generalization* .

### 3.5 Orange Canvas

O Orange Canvas [Orange Data Mining, 2021] é uma ferramenta *Open Source* destinada para análise de dados que possui recursos para criar todo fluxo de trabalho de um projeto de *Data Mining* ou modelos preditivos sem necessidade de Código de programação.

A plataforma possui um conjunto de *software* baseado em componentes para *Machine Learning* e *Data Mining*, desenvolvido no laboratório de bioinformática na Faculdade de Ciência da Computação e Tecnologias da Universidade de Ljubljana, na Eslovénia, em conjunto com uma comunidade de apoio *open source*. A primeira versão da plataforma foi lançada em 1996, com o nome de ML, e tratava-se de um *framework* de *Machine Learning* escrito na linguagem C++. Em 1997 adicionaram-se vínculos da linguagem *Python* o que levou à criação do *framework* chamado Orange.

### 3.6 Ensemble Learning

O *Machine Learning* baseado em *Ensemble Learning* [Zhang et al., 2012] tem como principal objetivo utilizar mais de uma técnica de *Machine Learning*, ou seja, ele realiza o treinamento de duas ou mais técnicas de aprendizagem de máquina que agrega os melhores resultados. O resultado visa proporcionar uma melhor acurácia e principalmente um modelo onde a presença de erros diminuam e o seu desempenho e performance seja mais confiável.

Os modelos utilizados devem ser treinados para uma mesma tarefa, a ideia é utilizar mais de um modelo de predição de forma simples (*weak learner*), para gerar um modelo agrupado mais complexo (*strong learner*) que é a soma de suas partes [Dong et al., 2020].

Quanto ao uso de uma técnica *Machine Learning* e do método *Ensemble Learning,* o que diferenciam ambos é a qualidade do resultado, com mais de uma técnica de treinamento a acurácia se torna mais confiável, pois o seu desempenho é beneficiado quando comparado apenas com uma técnica, pois há uma redução significativa na variação do resultado.

## 4. RESULTADOS

Os resultados da acurácia do conjunto de dados DDoS NTP foram realizados com a ferramenta Orange e com a utilização de três técnicas, sendo o KNN, SVM e o *Stacking*, que agrega os resultados do KNN e o SVM. A Figura 2 ilustra a ferramenta Orange e a funcionalidade da técnica *Stacking*.

'

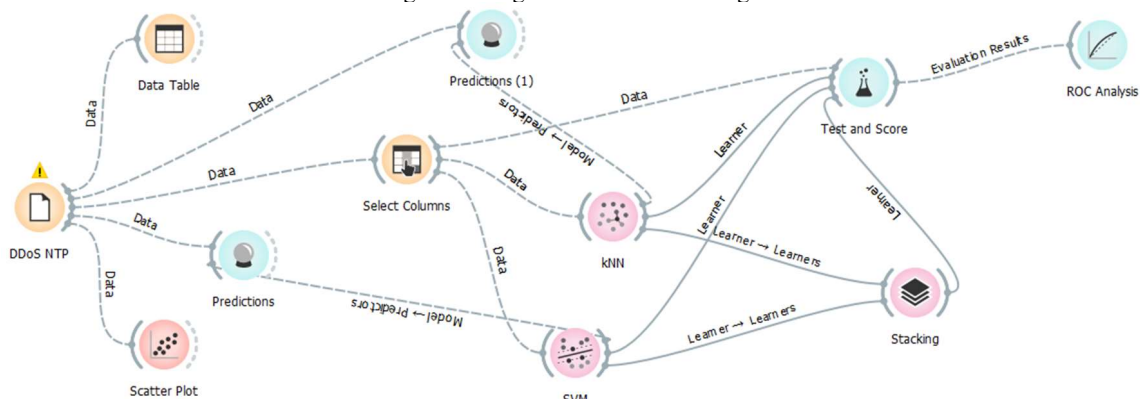

Figura 2: Diagrama ferramenta Orange

Fonte: [Orange Data Mining, 2021]

Os testes utilizaram 10% do tamanho total do *dataset* e com a variação dos valores nas Tabelas 1, 2 e 3 de teste e treinamento de cada método, os melhores resultados estão identificados em destaque.

Tabelas 1. Método KNN

| MODELO | AUC | CA | F1 | PRECISION | RECALL | TRAIN/TEST | TRAIN/SETSIZE |
|---|---|---|---|---|---|---|---|
| KNN | 0,984 | 0,997 | 0,997 | 0,997 | 0,997 | 5 | 10% |
| **KNN** | **0,985** | **0,998** | **0,998** | **0,998** | **0,998** | **10** | **10%** |
| KNN | 0,985 | 0,997 | 0,997 | 0,997 | 0,997 | 20 | 10% |

Tabelas 2. Método SVM

| MODELO | AUC | CA | F1 | PRECISION | RECALL | TRAIN/TEST | TRAIN SETSIZE |
|---|---|---|---|---|---|---|---|
| SVM | 0,436 | 0,793 | 0,874 | 0,977 | 0,793 | 5 | 10% |
| **SVM** | **0,470** | **0,825** | **0,894** | **0,978** | **0,825** | **10** | **10%** |
| SVM | 0,456 | 0,775 | 0,863 | 0,977 | 0,775 | 20 | 10% |

Tabelas 3. Método Stacking

| MODELO | AUC | CA | F1 | PRECISION | RECALL | TRAIN/TEST | TRAIN/SETSIZE |
|---|---|---|---|---|---|---|---|
| KNN | 0,98441 | 0,99748 | 0,99744 | 0,99742 | 0,99748 | 5 | 10% |
| STACK | 0,98344 | 0,99739 | 0,99734 | 0,99732 | 0,99739 | 5 | 10% |
| SVM | 0,43053 | 0,79327 | 0,87423 | 0,97719 | 0,79327 | 5 | 10% |
| **KNN** | **0,98467** | **0,99756** | **0,99755** | **0,99755** | **0,99756** | **10** | **10%** |
| **STACK** | **0,98428** | **0,99741** | **0,99738** | **0,99736** | **0,99741** | **10** | **10%** |
| **SVM** | **0,47416** | **0,82524** | **0,89365** | **0,97755** | **0,82524** | **10** | **10%** |
| KNN | 0,98505 | 0,99747 | 0,99742 | 0,99741 | 0,99747 | 20 | 10% |
| STACK | 0,98440 | 0,99737 | 0,99731 | 0,99730 | 0,99737 | 20 | 10% |
| SVM | 0,46035 | 0,77470 | 0,86262 | 0,97741 | 0,77470 | 20 | 10% |

## 5. DISCUSSÃO E CONCLUSÃO

As técnicas de *Machine Learning* a cada dia estão oferecendo uma maior diversidade de métodos e ferramentas em diferentes áreas, principalmente de detecção de invasões e segurança de redes.

Neste proposito, o tema foi estudado e observado utilizando os métodos kNN, SVM juntamente a técnica de *Stacking,* tornando-se possível o estudo do conjunto de dados com diversos parâmetros e técnicas.

Assim, o resultado tornou-se confiável e apresentou diversas possibilidades diferentes de estudos, principalmente observando as reações das técnicas de invasões.

'

Também observou-se os resultados de cada técnica, verificando que os melhores índices foram obtidos com a utilização de 10% do tamanho do conjunto de dados, e o uso de treino e teste que a ferramenta Orange oferece com os parâmetros fixados no valor 10. A métrica de Acurácia precisa ser observada, pois é ela que define se a técnica e os índices foram satisfatórios.

Os resultados mostraram satisfatórios, com a Acurácia da técnica *Ensemble Learning* atingindo o valor de **0,984,** e o de kNN **0,985**.

Por fim, o objetivo deste trabalho foi o estudo da técnica de *Ensemble Learning* aplicando a técnica de *Stacking* na ferramenta Orange para identicação de DDoS. Isto possibilitou observar e compreender que os resultados obtidos atingiram níveis elevados, das técnicas kNN e SVM e juntamente com a técnica de *Ensemble Learning,* que agrega os resultados das demais técnicas. Com isso, os valores obtidos tornaram-se mais fundamentados, podendo ser utilizados por pesquisadores em trabalhos futuros.

# 6. REFERÊNCIAS

Arshi, M., Md Nasreen, and Karanam Madhavi, 2020. "A Survey of DDOS Attacks Using Machine Learning Techniques." E3S Web of Conferences 184: 01052. doi:10.1051/e3sconf/202018401052

Buczak, A. L. e Guven, E., 2016. A survey of data mining and machine learning methods for cyber security intrusion detection. IEEE Communications Surveys Tutorials, 18(2):1153–1176.

Chand, N., Mishra, P., Krishna, C. R., Pilli, E. S., & Govil, M. C. (2016, April). A comparative analysis of SVM and its stacking with other classification algorithm for intrusion detection. In *2016 International Conference on Advances in Computing, Communication, & Automation (ICACCA)(Spring)* (pp. 1-6). IEEE.

Cortes, O. A. C., & de Oliveira Melo, W. E. (2021). Utilizando Análise de Sentimentos e SVM na Classificação de Tweets Depressivos. *Anais do Computer on the Beach*, *12*, 102-110.

Cover, T., & Hart, P. (1967). Nearest neighbor pattern classification. *IEEE transactions on information theory*, 21-27.

DDoS Evaluation Dataset (CIC-DDoS2019)" University of New Brunswick Est.1785. Accessed December 07, 2021. https://www.unb.ca/cic/datasets/ddos-2019.html.

Dong, X., Yu, Z., Cao, W., Shi, Y., & Ma, Q. (2020). A survey on ensemble learning. *Frontiers of Computer Science*, *14*(2), 241-258.

Ennaji, S., El Akkad, N., & Haddouch, K. (2021, October). A Powerful Ensemble Learning Approach for Improving Network Intrusion Detection System (NIDS). In *2021 Fifth International Conference On Intelligent Computing in Data Sciences (ICDS)* (pp. 1-6). IEEE.V. Anuja Kumari, R.Chitra , "Classification Of Diabetes Disease Using Support Vector Machine."International Journal of Engineering Research and Applications

Faria, M. M. (2016). Detecção de intrusões em redes de computadores com base nos algoritmos KNN, K-Means++ e J48. *São Paulo: Dissertação (Programa de Mestrado em Ciência da Computação)—Faculdade Campo Limpo Paulista–FACCAMP*.

Guan, Y. e Ge, X., 2018. "Distributed attack detection and secure estimation of networked cyber-physical systems against false data injection attacks and jamming attacks". IEEE Transactions on Signal and Information Processing over Networks, 4(1):48–59.

Orange Data Mining. Accessed December 20, 2021. https://orangedatamining.com/

Othman, S. M., Alsohybe, N. T., Ba-Alwi, F. M., & Zahary, A. T. (2018). Survey on intrusion detection system types. *International Journal of Cyber-Security and Digital Forensics*, *7*(4), 444-463.

Saini, Parvinder Singh, Sunny Behal, and Sajal Bhatia, 2020. "Detection of DDoS Attacks Using Machine Learning Algorithms." 7th International Conference on Computing for Sustainable Global Development (INDIACom), 2020. doi:10.23919/indiacom49435.2020.9083716.

Sharafaldin, I., Lashkari, A. H., Hakak, S., & Ghorbani, A. A. (2019, October). Developing realistic distributed denial of service (DDoS) attack dataset and taxonomy. In *2019 International Carnahan Conference on Security Technology (ICCST)* (pp. 1-8). IEEE.

Turcato, Afonso Celso, 2020. "Desenvolvimento de método para detecção de intrusão em redes PROFINET baseado em técnicas de Aprendizado de Máquina". Disponível em: https://www.teses.usp.br/teses/disponiveis/18/18153/tde-16072021- 172236/publico/TeseTurcatoAfonsoCelsoCorrig.pdf.

Yadav, S., & Subramanian, S. (2016, March). Detection of Application Layer DDoS attack by feature learning using Stacked AutoEncoder. In *2016 international conference on computational techniques in information and communication technologies (icctict)* (pp. 361-366). IEEE.

Zhang, C., & Ma, Y. (Eds.). (2012). *Ensemble machine learning: methods and applications*. Springer Science & Business Media.

'